\documentclass[useAMS,usenatbib,usegraphicx]{mn2e}

\newcommand{\aips}{{\sc aips}}
\newcommand{\difmap}{{\sc difmap}}
\newcommand{\mus}{$\mu$Jy~beam$^{-1}$}
\newcommand{\mms}{mJy~beam$^{-1}$}
\newcommand{\lensclean}{{\sc LensClean}}  
\newcommand{\lensmem}{{\sc LensMEM}} 
\newcommand{\hunits}{km$\,$s$^{-1}$Mpc$^{-1}$}
\newcommand{\iraf}{{\sc IRAF}}
\newcommand{\multidrizzle}{{\it multidrizzle}}
\newcommand{\starlink}{{\it Starlink}}
\newcommand{\gaia}{{\sc gaia}}
\newcommand{\scimg}{{\sc scimg}}
\newcommand{\donewparg}{\smallskip \noindent}

\newcommand{\ds}{$D_{\rmn{s}}$}
\newcommand{\dds}{$D_{\rmn{ls}}$}

\title[CLASS B0631+519: Last of the CLASS lenses]{CLASS B0631+519: Last of the CLASS lenses}
\author[York et al.]{T. York$^{1}$\thanks{E-mail: tyork@jb.man.ac.uk (TY)}, N. Jackson$^{1}$, I. W. A. Browne$^{1}$, L. V. E. Koopmans$^{2}$, J. P. McKean$^{3}$,\newauthor
M. A. Norbury$^{1}$, A. D. Biggs$^{4}$, R. D. Blandford$^{5}$, A. G. de Bruyn$^{2,6}$,\newauthor
C. D. Fassnacht$^{2}$, S. T. Myers$^{7}$, T. J. Pearson$^{8}$, P. M. Phillips$^{1}$,\newauthor
A. C. S. Readhead$^{8}$, D. Rusin$^{9}$ and P.N. Wilkinson$^{1}$ \\
$^{1}$University of Manchester, Jodrell Bank Observatory, Macclesfield, Cheshire, SK11 9DL, UK\\
$^{2}$Kapteyn Astronomical Institute, Postbus 800, NL-9700 AV Groningen, the Netherlands\\
$^{3}$Department of Physics, University of California, Davis, CA 95616, USA\\
$^{4}$Joint Institute for VLBI in Europe, Postbus 2, 7990 AA Dwingeloo, the Netherlands\\
$^{5}$KIPAC, Stanford University, 2575 Sand Hill Road, Menlo Park, CA 94025, USA\\
$^{6}$ASTRON, Postbus 2, 7990 AA, Dwingeloo, the Netherlands\\
$^{7}$National Radio Astronomy Observatory, P.O. Box 0, Socorro, NM 87801, USA\\
$^{8}$California Institute of Technology, Pasadena, CA 91125, USA\\
$^{9}$Department of Physics and Astronomy, University of Pennsylvania, 209 South 33rd Street, Philadelphia, PA 19104, USA }

\begin{document}
\date{2004 September 15}

\pagerange{\pageref{firstpage}--\pageref{lastpage}} \pubyear{2005}

\maketitle

\label{firstpage}

\begin{abstract}
We report the discovery of the new gravitational lens system
CLASS B0631+519. Imaging with the VLA, MERLIN and the VLBA reveals 
a doubly-imaged flat-spectrum radio core, a doubly-imaged steep-spectrum
radio lobe and possible quadruply-imaged emission from a second lobe. The 
maximum separation between the lensed images is 1.16~arcsec. High resolution 
mapping with the VLBA at 5~GHz resolves the most magnified image of the radio 
core into a number of sub-components spread across approximately 20~mas.
No emission from the lensing galaxy or an odd image is 
detected down to 0.31~mJy (5$\sigma$) at 8.4~GHz. Optical and near-infrared
imaging with the ACS and NICMOS cameras on the HST show that there are
two galaxies along the line of sight to the lensed source, as previously
discovered by optical spectroscopy. We find that the foreground galaxy 
at $z$=0.0896 is a small irregular, and that the other, at $z$=0.6196 is a 
massive elliptical which appears to contribute the majority of the
lensing effect. The host galaxy of the lensed source is detected in the
HST near-infrared imaging as a set of arcs, which form a nearly complete
Einstein ring. Mass modelling using non-parametric techniques can 
reproduce the near-infrared observations and indicates that the small
irregular galaxy has a (localised) effect on the flux density distribution
in the Einstein ring at the 5-10\% level.

\end{abstract}

\begin{keywords}
gravitational lensing; cosmology.
\end{keywords}

\section{Introduction}

Strong gravitational lensing occurs when a background source is multiply 
imaged by the gravitational field of an intervening massive galaxy or 
cluster of galaxies. Observations of gravitational lenses have been used 
in many astrophysical applications, such as in modelling the mass distributions 
of distant galaxies and as constraints on cosmological models \citep{KSW2004}. 
CLASS (Cosmic Lens All-Sky Survey; \citealt{class1,class2}) and JVAS
(Jodrell Bank - VLA Astrometric Survey; \citealt{jvas1,jvas2,jvas3})
are radio surveys of the northern sky designed to find gravitationally
lensed compact radio sources. In JVAS and CLASS 11685 sources in a statistically 
well-defined sample have been examined and 22 lens systems found. In this paper, 
we announce the discovery of the last of these systems, CLASS B0631+519. This lens
system exhibits complex radio structure over scales from 3.6~mas to 1.16~arcsec
and has a nearly complete infrared Einstein ring, similar to that in B1938+666
\citep{jvas1,1938a,1938b}. It is a system with one of the 
richest lensed image structures known and is thus an ideal system with
which to probe mass properties of the lensing galaxy. B0631+519 is a member of
the CLASS statistically complete sample.

\donewparg In Section \ref{SecRadObs} we discuss radio observations of B0631+519
made using the Very Large Array (VLA), Very Long Baseline Array (VLBA) and the
Multi-Element Radio-Linked Interferometer Network (MERLIN; \citealt{MERLIN}). Section
\ref{SecOptObs} presents optical and near-infrared images taken with
the William Herschel Telescope (WHT) and the {\it Hubble Space
Telescope} (HST). In Section \ref{SecModel} we consider the data in
the context of the lens hypothesis and test some simple lens
models, and in Section \ref{SecConclude} we conclude by discussing the
future priorities for work on this lens.

\section{Radio observations}
\label{SecRadObs}

As described in \citet{class1} and \citet{class2}, potential lens candidates from CLASS
are tested by observing them with a range of angular resolutions and at several different 
frequencies. If the supposed multiple images in a lens candidate are found to have very 
different surface brightnesses at any angular resolution the lens hypothesis can usually
be rejected immediately, since lensing preserves surface brightness. In CLASS, candidates 
discovered with the VLA are observed with higher resolution using 
MERLIN, and, if not eliminated at this stage, are re-observed with
even higher resolution using the VLBA. Accordingly, radio observations
of B0631+519 have been made over frequencies ranging from 1.7~GHz to
15~GHz, with angular resolutions from 3.6~mas to 236~mas. The
observations are summarised in Table \ref{RadObs}.

\setcounter{table}{0}
\begin{table*}
\begin{center}
\begin{tabular}{llllll}
\hline
Date observed & Instrument & Frequency [GHz] & Integration time & Map residual noise [\mus] & Mean beam size [mas] \\
\hline
1994 March 05 & VLA & 8.4 & 16 sec & 410 & 234 \\
1999 July 02 & VLA & 15 & 354 sec & 360 & 142 \\
1999 August 03 & VLA & 8.4 & 510 sec & 86 & 236 \\
1999 December 10 & VLBA & 1.7 & 1.4 hr & 130 & 8 \\
2001 March 21 & VLBA & 5 & 1.7 hr & 82 & 3.6 \\
2001 July 07 & MERLIN & 5 & 19 hr & 120 & 52 \\
2002 December 01 & VLBA & 1.7 & 7.8 hr & 64 & 9 \\ 
2003 March 01 & MERLIN & 1.7 & 20 hr & 62 & 173 \\
\hline
\end{tabular}
\caption{Radio observations made of B0631+519 to date.}
\label{RadObs}
\end{center}
\end{table*}

\subsection{VLA}

B0631+519 was observed at 8.4~GHz on 1994 March 05 and 1999 August 03. The 1999 
observation was a follow-up to the 1994 snapshot observation and had almost 5 
times its sensitivity (Table \ref{RadObs}). The snapshot was made using a bandwith
of 50~MHz, and 3C48 was used to set the flux density scale; the follow-up observation
used a bandwidth of 100~MHz, and the flux density scale was set using 3C147.

\donewparg All VLA observations of B0631+519 were made with the VLA in `A' configuration, 
and were calibrated using \aips\footnote{\aips: NRAO's Astronomical Image 
Processing System}\ and mapped with the Caltech difference mapping program \difmap\  
\citep{DIFMAP}. Model component positions and flux densities from the VLA observations 
are displayed in Table \ref{PosFluxVLA}. 

\donewparg The deconvolved map from the 1999 follow-up observation is shown in
Figure \ref{VLAlong}. We label the detected radio components as shown in Figure \ref{VLAlong}. 
Components A1 and B are separated by 1.15~arcsec at a position angle (from brightest to 
dimmest source) of $+$135$\degr$, and the A1:B flux density ratio is 8:1. Components 
A1, A2 and B are unresolved. Component X is a knot of elevated surface brightness embedded in the jet- 
or arc-like feature that branches to the south of A2. There is some extended emission 
visible on the opposite side of A1 from X, which we label Y, although the detection is 
marginal (about 2.3$\sigma$). The total flux density in the model is 42$\pm$2~mJy
(where we have assumed 5\% errors on absolute flux density measurements from the VLA at 8.4~GHz).  

\donewparg In the snapshot A1, A2 and B were detected, while X and Y were not. The snapshot shows an
A1:B flux density ratio of about 7:1, and a total flux density from the candidate of 
46$\pm$2~mJy. This is 10\% higher than the equivalent quantity in the follow-up map.
The discrepancy might be accounted for by uncertainty in the absolute amplitude calibration 
between the 1994 and the 1999 observations, but it could also indicate that the source is variable. 
The mean size of the synthesized beam varies by less than 3~mas between the two (uniformly 
weighted) maps, or by 2\% in area, so it is unlikely that 10\% of the emission has been resolved 
out in one map with respect to the other. The A1:B flux density ratio also changes by 16\%
between the two observations, which is hard to reconcile with a global mis-calibration of 
the amplitude scale. The signal-to-noise ratios on A1 and B were at least 10 in all VLA 
observations.

\setcounter{figure}{0}
\begin{figure}
\includegraphics[width=8cm]{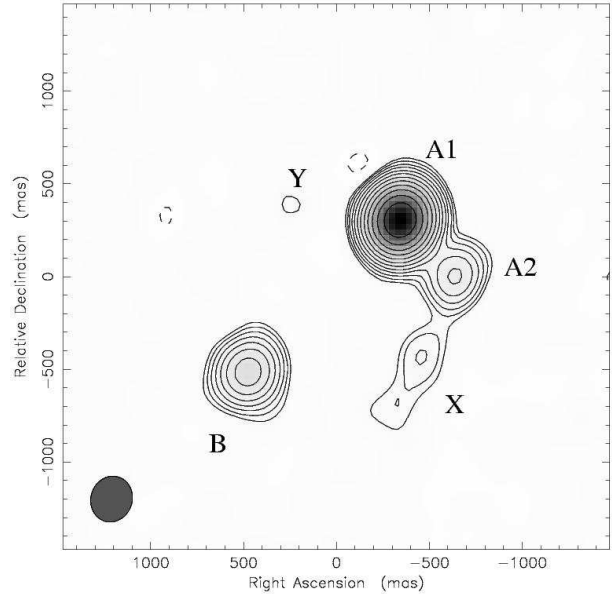}
\caption{Radio map of the VLA 8.4~GHz data taken on 1999 August 03. 
The RMS noise in the residual map is 86~\mus. The data are naturally weighted. 
The synthesized beam is sized 250$\times$223~mas with the major axis oriented 
along a position angle (P.A.) of $-$17.5$\degr$. The lowest contours are 
plotted at -3 and 3 times the RMS noise level; successive contours represent 
an multiplicative increase in surface brightness by a factor of 1.5, up to the 
highest contour level of 22.8~\mms.  Component A1 is located at RA: 06$^h$ 
35$^m$ 12.31402$^s$, Dec: $+$51$\degr$ 57' 01.8026\arcsec (J2000.0) as measured 
by the VLBA.}
\label{VLAlong} 
\end{figure}

\donewparg A further observation of B0631+519 was made with the VLA at 15~GHz
on 1999 July 02 for 354~seconds, with a bandwidth of 100~MHz. The amplitude 
scale was set with observations of 3C286. The deconvolved, naturally weighted 
map is shown in Figure \ref{VLA15}.
The total flux density detected at 15~GHz is 34$\pm3$~mJy, and the A1:B flux density 
ratio is 7.5:1. Both A1 and B are unresolved by the VLA at this frequency. A2 
is not detected, suggesting that it has a steeper spectral index than A1 and B. 

\setcounter{figure}{1}
\begin{figure}
\includegraphics[width=8cm]{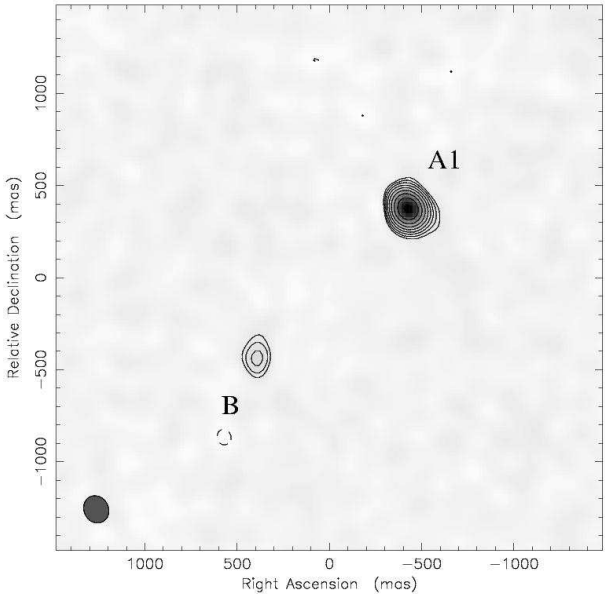}
\caption{15~GHz VLA data taken on 1999 July 02. The synthesized beam 
is 152$\times$132~mas oriented with the major axis at a P.A. of $+$26$\degr$, 
and the RMS noise in the residuals is 360~\mus. The data are naturally 
weighted. Contours are plotted as for Figure 1; the highest contour 
value is 27.8~\mms. }
\label{VLA15} 
\end{figure}

\setcounter{table}{1}
\begin{table*}
\begin{tabular}{lllll}
\hline
Component & Frequency [GHz] & \multicolumn{2}{c}{Position offsets [mas]} & Flux density [mJy] \\
& & $\Delta$R.A. & $\Delta$Dec. & \\
\hline
A1 & 8.4 & $+$0$\pm$1 & $+$0$\pm$1 & 34.3$\pm$1.7  \\
   & 15   & $+$0$\pm$2 & $+$0$\pm$2 & 29.6$\pm$3 \\
A2 & 8.4 & $-$297$\pm$10 & $-$300$\pm$10 & 2.2$\pm$0.1  \\
   & 15   & - & - & $<$1.8 \\
B  & 8.4 & $+$819$\pm$5 & $-$819$\pm$5 & 4.2$\pm$0.2  \\
   & 15   & $+$813$\pm$19  & $-$805$\pm$19 & 3.95$\pm$0.4  \\
X  & 8.4 & $-$114$\pm$33  & $-$754$\pm$33  &  0.98$\pm$0.1 \\
   & 15   & - & - & $<$1.8 \\
Y  & 8.4 & $+$577$\pm$70  & $+$81$\pm$70  & 0.2$\pm$0.1  \\
   & 15   & - & - & $<$1.8 \\
\hline
\end{tabular}
\caption{Model components from radio observations of B0631+519 with the VLA. 
All position offsets are relative to A1. The 8.4~GHz flux densities and 
positions are derived from the 1999 follow-up data, not the 1994 discovery data. 
Non-detections are quoted at the 5$\sigma$ level throughout this paper.
Errors on positions of Gaussian model components are estimated by 
dividing the beam size by the peak signal-to-noise ratio (SNR). 
Errors on flux densities are assumed to be subject to a 5\% random 
error in absolute flux density calibration at 8.4~GHz and a 10\% error at 15~GHz.}
\label{PosFluxVLA}
\end{table*}

\subsection{MERLIN}

B0631+519 was observed at 5~GHz using MERLIN on 2001 July 07, over a
period of 19 hours. Telescope gains were calibrated using the source OQ208, and
the observations were phase referenced to the source JVAS J0642+5247. 
All MERLIN data were reduced by initial amplitude calibration and flagging with
the MERLIN D-programs, followed by phase and amplitude calibration and further 
flagging in \aips. \difmap\ was used to map and self-calibrate the data. 3C286
was used to set the flux density scale for all MERLIN observations. 
Model component positions and flux densities from the MERLIN observations are 
shown in Table \ref{PosFluxMERLIN}. 

\donewparg The deconvolved 5~GHz map is shown in Figure \ref{MERLIN5}. Component B 
has been resolved into two components (B1 and B2). A1 and B1 appear compact, but 
A2 and B2 are slightly resolved. X and Y have been resolved out and there is no
sign of the extended emission revealed by the 8.4~GHz VLA observations.

\setcounter{figure}{2}
\begin{figure}
\includegraphics[width=8cm]{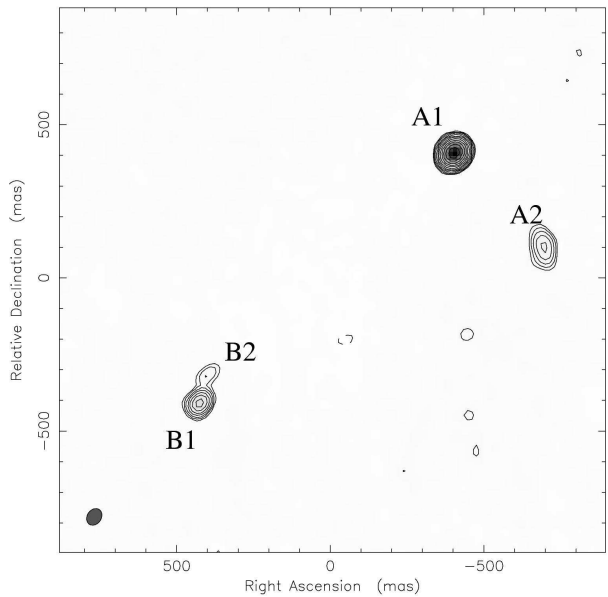}
\caption{Radio map of the 5~GHz MERLIN observations taken on 2001 July 07. 
The data were naturally weighted. The beam size is 58$\times$46~mas with 
the beam's major axis being oriented at a P.A. of $-$34$\degr$. The RMS 
noise in the residuals is 120~\mus. Contours are plotted at levels of 
(-3, 3, 4.5, 6.75, 10.1, 15.2, 22.8, 34.2, 51.3, 76.9, 115, 173, 259) 
times the RMS noise.}
\label{MERLIN5} 
\end{figure}

\donewparg Additional MERLIN observations were made at 1.7~GHz on 2003 March 01. 
Phase referencing was carried out using the source J0636+5009. 
B0552+398 was used to calibrate the telescope gains. The newly upgraded 
76~metre Lovell Telescope was included in the interferometer network, allowing an RMS 
noise level in the deconvolved map of 62~\mus to be reached in 20 hours. 
The deconvolved 1.7~GHz map is shown in Figure \ref{MERLIN1.6}. The map 
shows emission both to the west and east of image A1. We identify these regions of 
emission as low-frequency counterparts to components X and Y seen in the VLA data. B1 also 
develops an extension to the west, which we label Z. 

\setcounter{figure}{3}
\begin{figure}
\includegraphics[width=8cm]{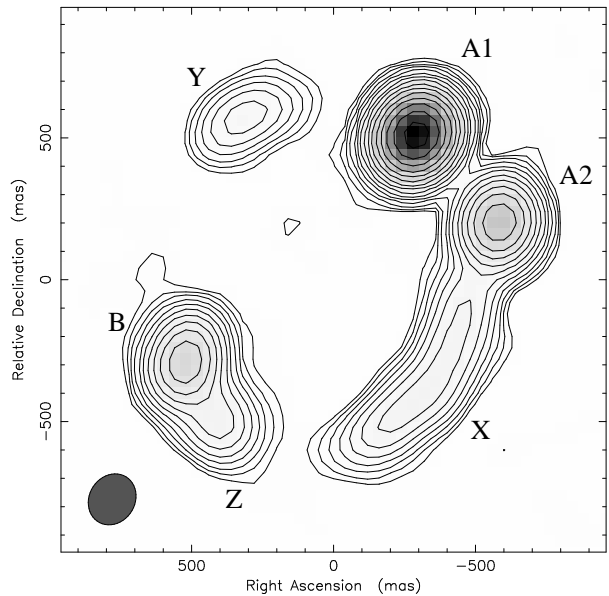}
\caption{Radio map of the 1.7~GHz MERLIN observations taken on 2003 March 01. 
The data were naturally weighted. The synthesized beam is sized 186$\times$161~mas 
at P.A. $-$31.7$\degr$. Contours are plotted at factors of (-3, 3, 4.5, 6.75, 
10.1, 15.2, 22.8, 34.2, 51.3, 76.9, 115, 173, 259, 389, 584, 876) times the 
RMS noise.}
\label{MERLIN1.6} 
\end{figure}

\setcounter{table}{2}
\begin{table*}
\begin{tabular}{lllll}
\hline
Component & Frequency [GHz] & \multicolumn{2}{c}{Position offsets [mas]} & Flux density [mJy] \\
& & $\Delta$R.A. & $\Delta$Dec. & \\
\hline
A1 & 1.7 & $+$0$\pm$1 & $+$0$\pm$1 & 66.4$\pm$6.6 \\
   & 5    & $+$0$\pm$1 & $+$0$\pm$1 & 46.9$\pm$4.7 \\
A2 & 1.7 & $-$293$\pm$1 & $-$313$\pm$1  & 19$\pm$1.9 \\
   & 5    & $-$293$\pm$3 & $-$308$\pm$3  & 4.2$\pm$0.4 \\
B  & 1.7 & $+$809$\pm$1 & $-$800$\pm$1  & 11.8$\pm$1.2 \\
B1 & 5    & $+$828$\pm$1 & $-$816$\pm$1  & 5.4$\pm$0.5 \\
B2 & 5    & $+$801$\pm$8 & $-$722$\pm$8  & 1.3$\pm$0.1 \\
X  & 1.7 & $-$40      &  $-$880        & 11$\pm$1 \\
   & 5    & -           & -            & $<$0.6 \\
Y  & 1.7 & $+$595$\pm$6 & $+$67$\pm$6   & 2.7$\pm$0.3 \\
   & 5    & -           & -            & $<$0.6 \\
Z  & 1.7 & $+$691$\pm$5 & $-$1016$\pm$5 & 5.0$\pm$0.5 \\
   & 5    & -           & -            & $<$0.6 \\
\hline
\end{tabular}
\caption{Radio components detected in B0631+519 with MERLIN. All 
position offsets are relative to A1. X is a large arc that is not 
well described by a single gaussian component; the position quoted 
is an estimate for the centre of the feature. The integrated flux 
density of X was derived by summing within the 3$\sigma$ contour.
We have assumed a 10\% absolute calibration error on flux densities 
from MERLIN. Position errors are estimated from the 
beam size divided by the peak SNR.}
\label{PosFluxMERLIN}
\end{table*}

\subsection{VLBA}

The VLBA was used to observe B0631+519 at 5~GHz on 2001 March 21 over a 
period of 1.7 hours. Data were taken using four contiguous 
8~MHz bands and a single hand of circular polarisation. The observations were 
phase-referenced using the calibrator source J0631+5311 and employed 2~bit 
sampling during digitisation; correlation was performed using the VLBA correlator 
in Socorro. During correlation each band was subdivided into 16 0.5~MHz channels 
and the data were averaged into 2~s integrations. 

\donewparg All VLBA observations were reduced using \aips\ to flag and calibrate the data, 
following standard recipes for the VLBA. The \aips\ task \scimg\ was then used 
to map the data through cycles of CLEAN \citep{CLEANcite} followed by self-calibration.
Flux densities and positions of VLBA components are shown in Table \ref{PosFluxVLBA}. 

\donewparg The deconvolved 5~GHz maps are shown in Figure \ref{VLBA5}. Uniform weighting was 
employed. The resolution achieved was 3.6~mas, sufficient to split A1 into four 
sub-components (labelled A1a to A1d) spread over an angle of about 20~mas. 
B1 remained unresolved, and no emission from the A2, B2, X, Y or Z components was 
detected down to the 5$\sigma$ limit of 390~\mus, indicating that these components 
have probably been resolved out. 

\donewparg We note that the sum of the flux densities of components A1a to A1d 
(46$\pm$2~mJy) is consistent with the MERLIN 5~GHz flux of component A1 to 
within the measurement errors, which implies both that A1 did not vary by 
more than 5\% over the 4 months separating the observations and also that no radio
structure seen in A1 by MERLIN was resolved out by the VLBA, despite the 
factor of $\sim$14 increase in resolution. However, another possible 
explanation is that an increase in source flux density (due to intrinsic variability)
cancelled out any loss of emission due to resolution effects. The same conclusions 
and caveats apply to the MERLIN and VLBA observations of B1 at 5~GHz. 

\donewparg Comparison of the total A1 flux density to the flux density of B1 suggests an A1:B1 
magnification ratio of $\sim$9:1, and hence that sub-components in B1 corresponding 
to those in A1 would span about 2~mas. Resolving the expected sub-components in B1 
is thus well within the capabilities of VLBI (Very Long Baseline Interferometry).

\setcounter{figure}{4}
\begin{figure}
\includegraphics[width=7cm]{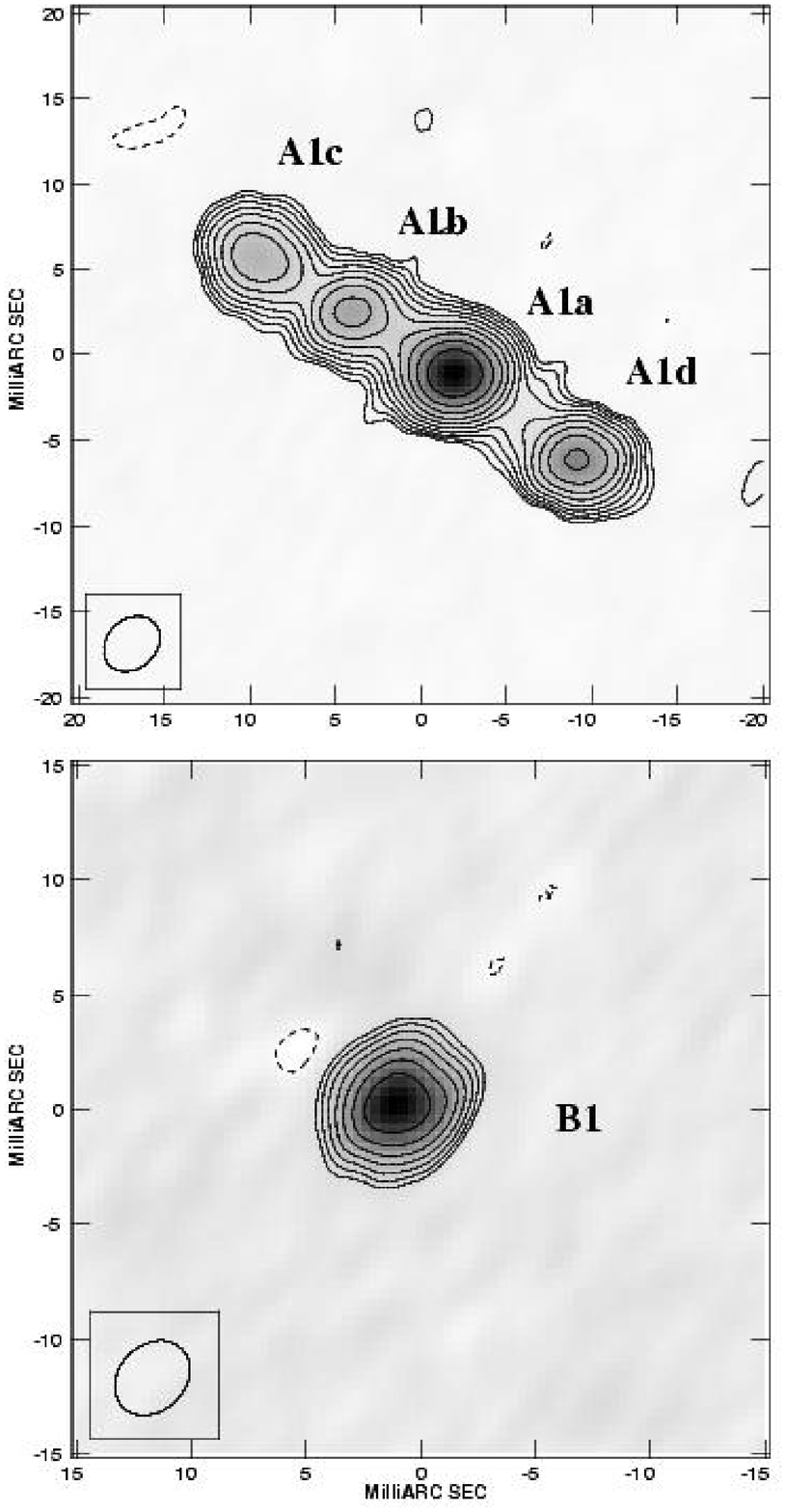}
\caption{Radio map of the 5~GHz VLBA observations taken on 2001 March 21, 
showing image A (top) and image B (bottom). The data were uniformly 
weighted. The beam-size is 3.8$\times$3.4~mas oriented at P.A. 
$+$144$\degr$ and the RMS noise is 82~\mus. Contours are plotted at levels 
of (-3, 3, 4.5, 6.75, 10.1, 15.2, 22.8, 34.2, 51.3) times the RMS noise.}
\label{VLBA5} 
\end{figure}

\donewparg Two epochs of phase-referenced VLBA data were obtained at 1.7~GHz, 
on 1999 December 10 and 2002 December 01. The phase reference source used was
J0631+5311. The 1999 data was taken in two polarisations via 4 contiguous bands, 2 
bands to each polarisation. 2-bit sampling was used, and correlation at Socorro 
again subdivided each band into 16 0.5~MHz wide channels and averaged the data 
into 2~s integrations. The phase centre was taken halfway between images A1 and 
B1 to minimise bandwidth smearing. 

\donewparg The CLEANed, naturally weighted maps of A1, A2 and B1/B2 from the 2002 epoch are shown in Figure \ref{VLBA1.6}. 
A1 appears resolved but (unsurprisingly) smoothed compared to the 5~GHz VLBA map, while B1 
remains unresolved. A2 is resolved into an arc-like feature of low surface brightness that extends over
80~mas. B2 is partially resolved into an extended source. No emission was detected from components X, Y or Z. 

\donewparg The maps from the two 1.7~GHz epochs look very similar. The flux densities of A1, A2 and B2 
agree between the two epochs to within the measurement errors. The flux density of B1 decreases by 
about 9\% between 1999 and 2002; this decrease is not quite significant at the 2$\sigma$ level 
given the measurement errors of about 5\% for B1's flux density. The measurements of the A1/B1 image separation 
from the two epochs are consistent to within 0.5~mas, while the A1-B2 separation measurements agree to within 2~mas.

\donewparg To identify which of the four VLBA sub-components of A1 correspond to the flat-spectrum radio core in
the lensed source, we registered the 1.7~GHz map (from the 2002 epoch) with the 5~GHz map by using the position of
component B1 as a reference (since B1 is largely unresolved in both maps). To overlay B1 in this way required the 
1.7~GHz image to be translated by 1.7~mas south and 8.9~mas east because slightly different positions were used for 
the phase reference source J0631+5311 between the 5~GHz and 1.7~GHz observations. The translated 1.7~GHz map is shown overlaid on
the 5~GHz map in Figure \ref{VLBAOverlay}. Sub-component A1a appears to be the flat spectrum core, while the peak
in the 1.7~GHz map corresponds to component A1c. Therefore at 5~GHz we apparently see both approaching and receding
jets in the source AGN. 

\setcounter{figure}{5}
\begin{figure}
\includegraphics[width=7cm]{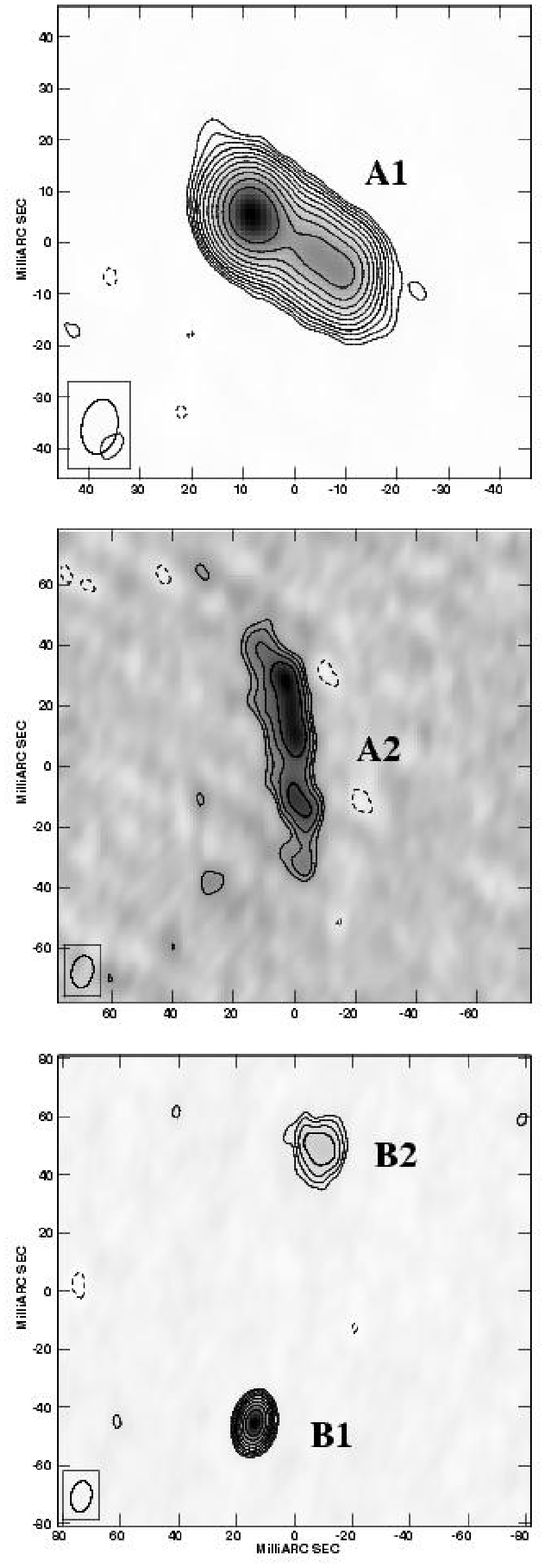}
\caption{Radio maps of the 1.7~GHz VLBA observations taken on 2002 December 01. 
The data were naturally weighted. The synthesized beam is an elliptical gaussian 
sized 11.2$\times$7.5~mas oriented with its major axis at a P.A. of $+$177$\degr$. 
The RMS noise is 64~\mus. Contours are plotted at levels of (-3, 3, 4.5, 6.75, 
10.1, 15.2, 22.8, 34.2, 51.3, 76.9, 115, 173) times the RMS noise.}
\label{VLBA1.6} 
\end{figure}

\setcounter{figure}{6}
\begin{figure}
\includegraphics[width=7cm]{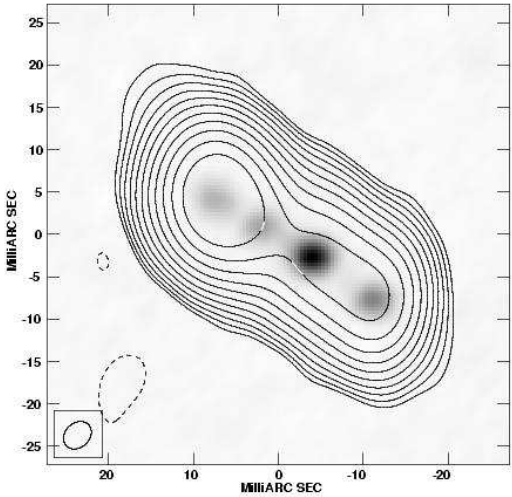}
\caption{The 1.7~GHz VLBA observations from 2002 December 01 plotted with the 
5~GHz data from the same instrument. The 5~GHz data are shown in greyscale,
while the 1.7~GHz data are shown contoured (with the same contour levels as in 
Figure \ref{VLBA1.6}).}
\label{VLBAOverlay} 
\end{figure}

\setcounter{table}{3}
\begin{table*}
\begin{tabular}{lllll}
\hline
Component & Frequency [GHz] & \multicolumn{2}{c}{Position offsets [mas]} & Flux density [mJy] \\
& & $\Delta$R.A. & $\Delta$Dec. & \\
\hline
A1  & 1.7 & $+$0$\pm$0.1    & $+$0$\pm$0.1 & 51.4$\pm$3 \\ 
A1a & 5   & $+$0$\pm$0.1    & $+$0$\pm$0.1 & 20.5$\pm$1 \\ 
A1b & 5   & $+$5.7$\pm$0.1   & $+$3.5$\pm$0.1 & 8.8$\pm$0.4 \\ 
A1c & 5   & $+$11.1$\pm$0.1  & $+$6.6$\pm$0.1 & 7.3$\pm$0.4 \\ 
A1d & 5   & $-$7.2$\pm$0.1   & $-$5.0$\pm$0.1 & 9.2$\pm$0.5 \\ 
A2  & 1.7 & $-$300   & $-$306 & 6.6$\pm$0.7 \\  
    & 5   & -             & -           & $<$0.39 \\  
B1  & 1.7 & $+$812.5$\pm$0.1 & $-$827.5$\pm$0.1 & 5.6$\pm$0.3 \\  
    & 5   & $+$822.5$\pm$0.1 & $-$821.5$\pm$0.1 & 5.1$\pm$0.3 \\  
B2  & 1.7 & $+$790.7$\pm$0.4   & $-$732.9$\pm$0.5   & 2.2$\pm$0.2 \\  
    & 5   & -             & -             & $<$0.39 \\
\hline
\end{tabular}
\caption{Model components for the radio observations of B0631+519 with the VLBA. 
Stated errors on flux densities assume a 5\% random error in the absolute flux 
density scale calibration at the VLBA. 
The 1.7~GHz position offsets are given relative to the surface brightness peak in 
A1. For A2 and B2 at 1.7~GHz, we have doubled 
the error to reflect the uncertainty in the flux densities of these resolved 
components; their flux densities were obtained by summing within the 3$\sigma$ contour. 
Positions for A1, B1 and B2 were determined from the best-fit gaussians, restricted 
to the compact north-eastern feature in the case of A1 at 1.7~GHz. The 
position given for A2 is that of the brightest pixel. For the 5~GHz data, position 
offsets are taken relative to A1a at RA: 06$^h$ 35$^m$ 12.31390$^s$, Dec: +51$\degr$ 57' 01.7949\arcsec (J2000.0).}
\label{PosFluxVLBA}
\end{table*}

\subsection{Radio spectra from continuum data}

We plot the radio spectra for the model components as a function of frequency
in Figure \ref{SpecInd}. If the lens hypothesis is correct then all the lensed 
images of a source should have similar spectra, provided that the lens galaxy 
ISM does not scatter or absorb the source light, and that the source flux 
density is constant \citep{K&D2004}. A1 and B1 have similar spectra, as do 
A2 and B2. Possibly X, Y and Z also have similar spectra, although at 8.4~GHz
component Y is a marginal detection and component Z is a non-detection (we
have plotted a 5$\sigma$ upper limit). We suggest that there are three separate
source regions revealed by the radio observations, two of which are doubly-imaged
and one of which is quadruply imaged (generating images Y, Z and the arc X).
We will discuss the geometry of the source with respect to the lens caustics 
further in Section \ref{SecModel}.

\setcounter{figure}{7}
\begin{figure}
\includegraphics[width=9cm]{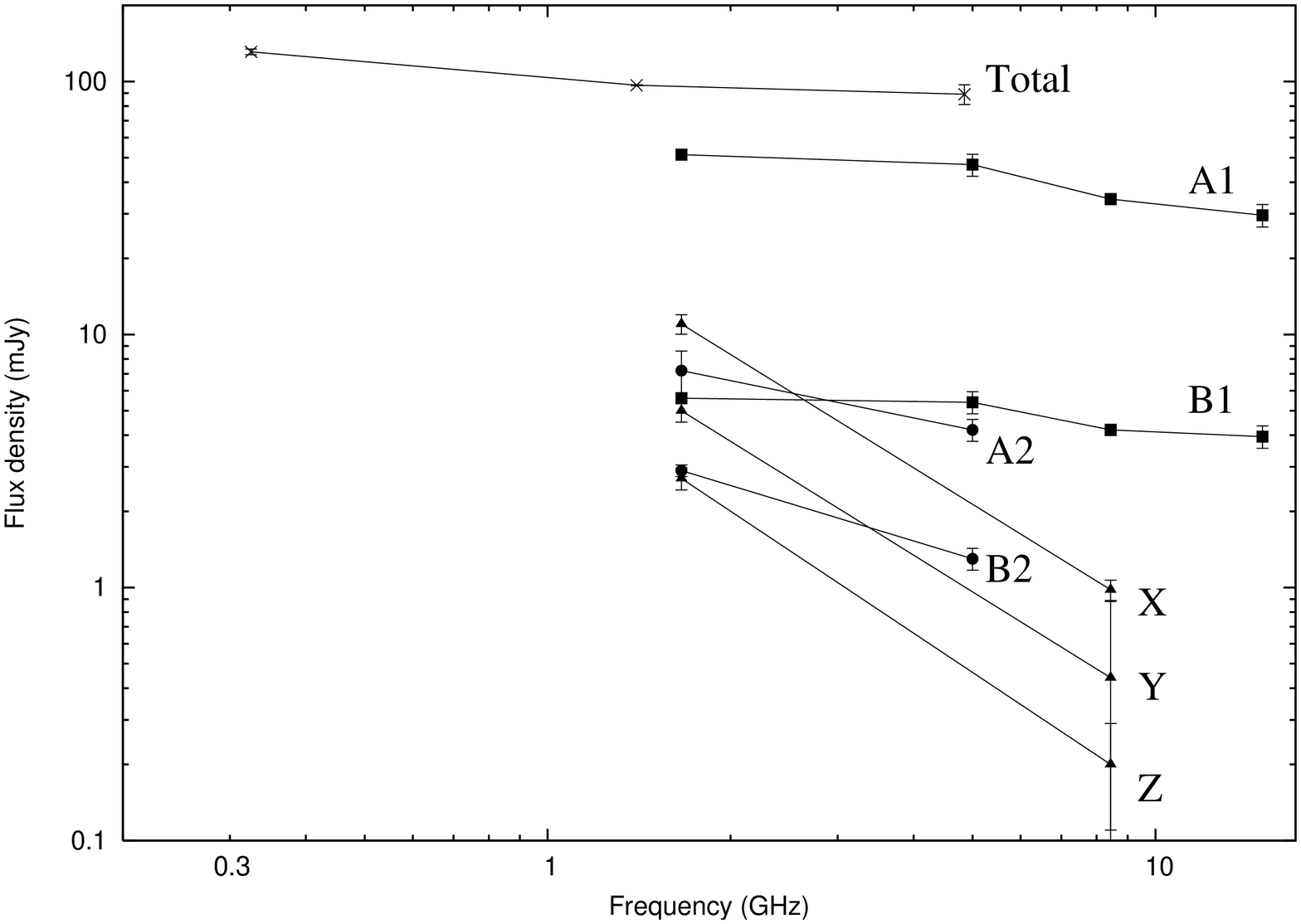}
\caption{Plot of the radio spectra of the various radio components identified in B0631+519. 
The total flux density is based on measurements at 327~MHz from WENSS \citep{WENSS}, at 1.4~GHz from NVSS \citep{NVSS} and at 5~GHz from the GB6 survey \citep{GB6}.}
\label{SpecInd} 
\end{figure}

\section{Optical observations}
\label{SecOptObs}

Typically lens galaxies are radio-quiet ellipticals, so optical/infrared imaging is necessary to establish the
lens galaxy position when the lenses are identified through radio searches. The position of the lens galaxy 
with respect to the lensed images can be used to constrain mass models, and in some systems unique 
information on the lensed source is available through optical imaging, for instance where an optical or infrared
Einstein ring is visible as seen in CLASS B1938+666 \citep{1938a,1938b}. We have observed 
B0631+519 at optical/near-infrared wavelengths with the WHT through an R-band filter and with the HST through F555W, 
F814W and F160W filters. Details of these observations are listed in Table \ref{OptObs}. 

\setcounter{table}{4}
\begin{table*}
\begin{center}
\begin{tabular}{llllllll}
\hline
Instrument & Detector & Date observed & Filter & Integration time [sec] & PSF FWHM [arcsec] & Plate scale [arcsec pixel$^{-1}$] \\
\hline
HST & ACS/WFC & 2003 August 19 & F555W & 2236 & 0.05 & 0.05 \\
HST & ACS/WFC & 2003 August 19 & F814W & 2446 & 0.08 & 0.05 \\
HST & NICMOS/NIC2 & 2004 February 15 & F160W & 2560 & 0.13 & 0.075 \\
WHT & Aux. port CCD & 2002 February 03 & R & 1800 & 0.7-0.9 & 0.108 \\
\hline
\end{tabular}
\caption{Optical and infrared observations made of B0631+519 to date.}
\label{OptObs}
\end{center}
\end{table*}

\subsection{WHT observations}

B0631+519 was observed with the 4.2~metre William Herschel Telescope
(WHT) on 2002 February 03, through an R-band filter. The detector used
was a 1024$\times$1024 pixel TEK CCD, having a total field of view of
111 $\times$ 111~arcsec. The data were calibrated using the \starlink\ 
package {\sc ccdpack}. B0631+519 was partially resolved, but there was 
insufficient resolution to separate any lensed images from the lens galaxy. 

\donewparg Aperture photometry was undertaken on the WHT image using 
observations of Landolt standard stars 93 407 and 100 280 \citep{Landolt} 
to correct for atmospheric extinction and to calibrate the brightness scale 
in magnitudes. An elliptical aperture with major diameter 4.3~arcsec and an 
axis ratio of 0.75 was used to obtain an R-band magnitude for B0631+519 of 
21.1$\pm$0.1, corrected for Galactic extinction using the dust maps of 
\citet{DustMaps}.

\subsection{HST optical and near-infrared observations}

B0631+519 was observed with the {\it Hubble Space Telescope} (HST) on
2003 August 19 and 2004 February 15 as part of HST proposal 9744, ``HST Imaging
of Gravitational Lenses'' (PI: C.S. Kochanek). Images were taken
using the Advanced Camera for Surveys (ACS) Wide Field Channel (WFC)
through the F555W and F814W filters. B0631+519 was also observed with
the NICMOS camera NIC2 through the F160W filter, in the MULTIACCUM 
observing mode. The wideband HST filters F555W, F814W and F160W 
approximately correspond to Johnson V, I and H respectively.
Details of the optical observations are shown in Table \ref{OptObs}. 
The ACS observations used a dither pattern consisting
of two pointings separated by 0.73~arcsec. Two exposures were taken at both pointings
to simplify cosmic ray rejection. The NICMOS data used a four-point dither
pattern that placed B0631+519 at the centre of each quadrant of NIC2
in turn. 

\donewparg The ACS and NICMOS data were debiased, dark-subtracted and 
flat-fielded using the OTFR (On-The-Fly-Recalibration) facility of the 
MAST HST archive. We then used the \multidrizzle\ script \citep{Mdrizzle} 
from within {\it PyRAF} to correct the ACS data for distortion, to eliminate 
cosmic rays and to combine dithered pointings into a single output frame, 
the shifts between frames being obtained from application of cross-correlation 
through the {\it crossdriz} task in \iraf. The NICMOS data were combined using 
{\it pydrizzle}  to interpolate the data onto an output image with a 
uniform pixel scale of 0.075~arcsec, using the standard distortion
solution for NIC2. Cosmic rays in the NICMOS data were identified and 
blanked by the OTFR pipeline itself. The \iraf\  task {\it pedsub}  was used 
to remove some residual dark signal that was evident in the output 
from the pipeline. Both the ACS and NICMOS images were aligned to the local 
celestial coordinate axes as part of the drizzling process.

\donewparg The HST images (Figure \ref{OpticalVIH}) reveal two galaxies at 
the lens system position, surrounded by arcs of emission from the lensed
source at longer wavelengths. We label these galaxies L1 and L2. L1 appears 
to be an elliptical galaxy, and is detected in all three bands. In the 
F814W and F160W images some emission from the source appears in
the form of arcs of extended emission encircling L1, consistent with the lens 
hypothesis if L1 is the primary lensing galaxy. L2 appears to be an irregular
galaxy and is seen in the F555W and F814W data, although in the latter it is confused
with one of the lensed arcs. L2 is not cleanly detected in the NICMOS image, and
may not be visible in the near-infrared.

\donewparg The ACS images have a large enough field of view to show the environment of B0631+519.
The nearest extended object is located 4.4~arcsec away at a P.A. of -72\degr, and 
has a total I$_{mag}$ $\sim$ 22.3 in a 4~arcsec diameter aperture. The object's light distribution 
is not smooth and is difficult to classify. 

\setcounter{figure}{8}
\begin{figure*}
\includegraphics[width=11cm]{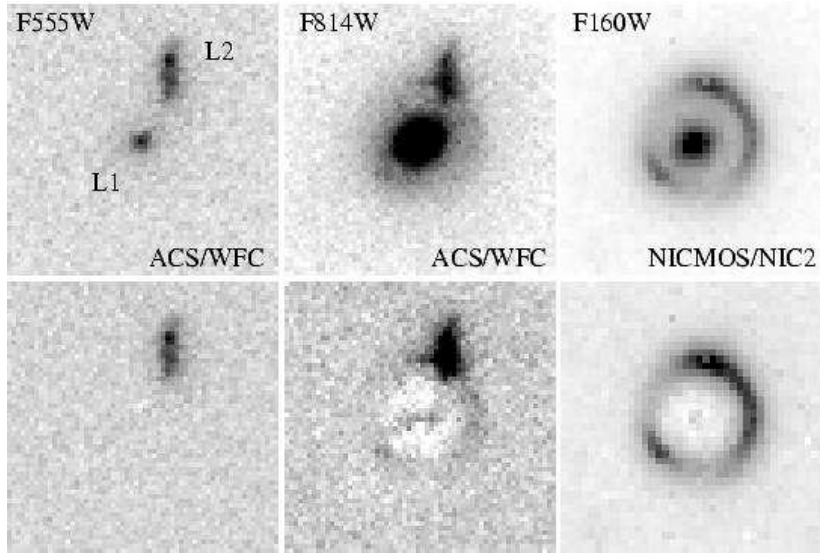}
\caption{HST images of B0631+519. The top row of images show the
lens before subtraction of L1, while the bottom row shows the residuals 
after subtraction of the best-fit S\'{e}rsic profile. North is up and
east is left. The images are 3$\times$3~arcsec in size, and
are centred on L1.}
\label{OpticalVIH} 
\end{figure*}

\donewparg We carried out surface-brightness fitting on L1 using a S\'{e}rsic 
light profile \citep{SersicProfile} convolved with appropriate PSFs 
produced by TinyTim \citep{KristTT}. The profile parameters were varied to 
minimise the sum of squared residuals between the data and the model. When 
fitting L1's surface-brightness profile we masked out L2 and the lensed images 
to avoid disturbing the fit. L2 and the lensed arcs do not have easily 
parametrizable surface light distributions, so we did not attempt to separate 
them in the F814W or F160W images, and report only the sum of 
light remaining after subtraction of L1. In the F555W image the arcs were not 
detected, so we used \gaia\ to fit an elliptical aperture to L2 (after
L1 had been subtracted). The centroid of L2 was found to be $-$0.33~arcsec east and
$+$0.74~arcsec north of L1. The axis ratio of the best-fit aperture was 0.56, with
the major axis having a position angle $-$1.1\degr east of north. Table \ref{OptMod} 
shows the properties of the S\'{e}rsic profiles subtracted from L1.

\setcounter{table}{5}
\begin{table}
\begin{center}
\begin{tabular}{llll}
\hline
Parameter & \multicolumn{3}{c}{Filter} \\
& F555W (V) & F814W (I) & F160W (H) \\ 
\hline
Offset in R.A. [mas] & $+$360$\pm$80 & $+$380$\pm$80 & $+$360$\pm$80 \\
Offset in Dec. [mas] & $-$500$\pm$80 & $-$530$\pm$80 & $-$500$\pm$80 \\
Total magnitude         & 22.7 & 20.0 & 17.7 \\
Effective radius [mas]  & 540 & 560 & 550 \\
Ellipticity             & 0.09 & 0.17 & 0.12 \\
Position angle          & $-$48 & $-$53 & $-$57 \\
Power-law index		& 0.29 & 0.28 & 0.30 \\
\hline
\end{tabular}
\caption{Parameters of the best-fit S\'{e}rsic profiles fitted to L1 in the three HST images. A power-law index
of 0.25 corresponds to a de Vaucouleurs surface-brightness profile. The total magnitudes are specified in
the Johnson photometric system, corrected for galactic extinction. Position offsets are given relative to
image A1.}
\label{OptMod}
\end{center}
\end{table}

\donewparg The ACS and NICMOS images following subtraction of L1 are shown in Figure 
\ref{OpticalVIH}. The S\'{e}rsic profile is a reasonable fit to the HST data, leaving
maximum residuals of 3\% and 4\% of L1's peak brightness in the case of the F814W
and F160W images respectively. The residuals are undetectable within the 
noise in the case of the F555W image, which has the lowest signal-to-noise 
ratio of all the HST observations. 

\setcounter{table}{6}
\begin{table}
\begin{center}
\begin{tabular}{llllllll}
\hline
Source & \multicolumn{4}{c}{Brightness} \\
& V & R & I & H \\
\hline
Total     & 22.1$\pm$0.1 & 21.1$\pm$0.1 & 19.9$\pm$0.1 & 17.3$\pm$0.1 \\
L1        & 22.8$\pm$0.1 &             & 20.0$\pm$0.1 & 17.8$\pm$0.1 \\
L2        & 23.0$\pm$0.1 &             &             & $>$23      \\
L2+images &              &             & 22.3$\pm$0.1 &             \\
Images    & $>$25      &             &             & 18.3$\pm$0.1 \\
\hline
\end{tabular}
\caption{Results of photometry on B0631+519. The figures reported in this table were 
measured within a circular aperture of 4~arcsec diameter. For the total luminosity 
of the subtracted S\'{e}rsic models see Table \ref{OptMod}. The WHT R-band
magnitude was calculated using a Kron elliptical aperture 4.3~arcsec in (major axis) diameter. All 
figures have been corrected for galactic and, in the case of the WHT, atmospheric extinction. Non-detections
are reported as lower limits on magnitude at the 5$\sigma$ level.}
\label{Photom}
\end{center}
\end{table}

\donewparg We carried out aperture photometry on B0631+519, using a circular aperture 
4~arcsec in diameter to measure the flux of the lens system as a whole, of the S\'{e}rsic model for L1
and of whatever light remained after subtraction of the S\'{e}rsic model from the data.
The measurements were converted into the Johnson magnitude system using
published zero points for HST magnitude systems. Galactic extinction was estimated from 
\citet{DustMaps}. The results are shown in Table \ref{Photom}. The differences in
L1 magnitudes between Tables \ref{OptMod} and \ref{Photom} are due to the fact that the 
values given in the latter are sums of the flux within finite apertures, while those in 
the former are equivalent to sums within infinite apertures.

\donewparg The lens galaxy position with respect to the lensed images, when known, 
provides two constraints on simple parametric lens models under the 
assumption that the centre of (smoothly distributed) light and the 
centre of mass for a galaxy are coincident. Since the lens galaxy in
B0631+519 is apparently radio-quiet, the radio and HST data must be 
overlaid as accurately as possible to establish the lens position. 
The lensed images at near-infrared and optical wavelengths exhibit no obvious 
point-like components that could be unambigiously aligned with the radio structure. 
We must therefore rely on absolute astrometric calibration based on stars visible 
in the field to align the optical and near-infrared images with radio data. 

\donewparg We used the 2MASS\footnote{See: http://pegasus.astro.umass.edu} catalogue with 
\starlink's \gaia\ to establish an astrometric 
calibration for the ACS images (F555W and F814W filters). 2MASS positions have no 
significant systematic offsets with respect to the radio-based ICRF \citep{2MASSoffset}.
The astrometric solutions were based on twelve star positions and had reported RMS 
errors of $\pm$0.08~arcsec in R.A. and declination, although examination
of objects around the field indicated positional agreement between the two calibrated
images to within 0.05~arcsec. In the case of NICMOS the relatively small 
field of view (19.5$\times$19.5~arcsec) meant that only one
star near to B0631+519 was imaged. The centre of this star was used 
as a reference to fix the absolute astrometry for the image. 
We estimate the astrometric error in the resulting solution for the 
drizzled NICMOS image at 0.08~arcsec, since the star's position on the
sky was derived from the astrometrically calibrated ACS F555W image. The
rotation of the image was checked by calculating the position angle of the 
vector from the field star to L1's position. This angle agreed with the
same quantity measured from the ACS images to within 0.05\degr.
Figure \ref{HST+Radio} shows the 1.7~GHz MERLIN map overlaid on the ACS 
and NICMOS images. 

\donewparg Fitting S\'{e}rsic profiles to L1 gave us best-fit pixel positions for the
galaxy centre in all three HST filters, which we converted
into celestial coordinates using the absolute astrometric solutions from 
the HST images. The sky position of image A1 (from the 1.7~GHz VLBA 
observations) was then subtracted to leave L1's offset on the sky from A1.
These offsets are reported in Table \ref{OptMod}. The agreement between
the offsets from the F555W and F160W images (which effectively share
the same astrometric solution) and the disagreement between F555W and F814W
suggests that the error in the galaxy position is dominated by the random
error in the astrometric solutions for F555W and F814W; error 
contributions from the galaxy subtraction procedure and from the distortion 
solutions for each camera are negligible by comparison. The mean offset 
from A1 to L1 (based on the ACS data only) is $+$0.37$\pm$0.08~arcsec in
R.A. and $-$0.52$\pm$0.08~arcsec in declination. 

\setcounter{figure}{9}
\begin{figure}
\includegraphics[width=7cm]{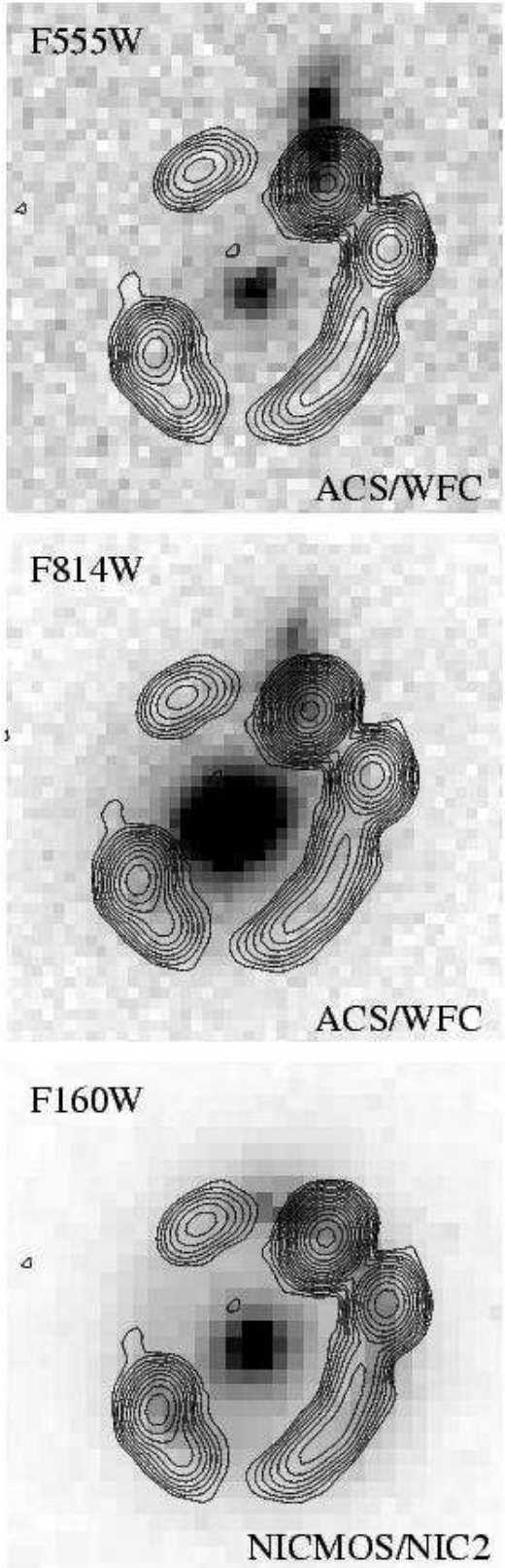}
\caption{The MERLIN 1.7~GHz radio data (contoured) overlaid on the 
NICMOS and ACS images (greyscale). The ACS images cover 2.45$\times$2.45 arcsec,
while the NICMOS image covers 2.475$\times$2.475 arcsec. All images are centred
on L1. North is up and east is left.}
\label{HST+Radio} 
\end{figure}

\subsection{Redshift-dependent properties of B0631+519}
\label{SecRedshifts}

\citet{JMK2004} have obtained an optical spectrum of B0631+519
with the W. M. Keck telescope. They report the discovery of 
two distinct galaxies at two different redshifts. The first is
an emission line galaxy at $z$=0.0896$\pm$0.0001; and the second,
which is the dominant source of emission in the spectrum, is an
elliptical galaxy at $z$=0.6196$\pm$0.0004. These spectral classifications
are consistent with the morphologies, colours and flux densities of the
two foreground galaxies detected in the HST imaging. Therefore, we 
associate the $z$=0.0896 emission line galaxy with L2, and the $z$=0.6196
elliptical with L1. \citet{JMK2004} refer to these galaxies in redshift 
order as G1 and G2 respectively; in this paper we have swapped the name 
order because the HST imaging demonstrates that L1 (G2) is the primary 
lensing galaxy, something that was not obvious from the spectroscopy alone.

\donewparg \citet{JMK2004} also discuss the tentative detection of a single
emission line from the lensed source at 5040\AA. Assuming that the 
line is either due to Ly$\alpha$ or Mg II emission, they suggest 
3.14 or 0.80 for the source redshift, respectively. However, further
spectroscopy will need to be carried out to test these possibilities.
We do not have enough (unambiguous) photometric information on the lensed 
images to attempt to choose between these possibilites by using photometric 
redshift techniques, but, in the absence of further spectroscopy, we can 
attempt to constrain the source redshift by determining the rest-frame luminosity 
$L$ of L1 and converting it into a velocity dispersion $\sigma$ using the 
fundamental plane of elliptical galaxies \citep{FP1,FP2}. This approach has been used
before by \citet{1938z} to estimate the source redshift in CLASS B1938+666.

\donewparg We converted the apparent magnitudes of L1 and L2 as measured by 
the HST (after translation into Johnson filters V, I and H, and following application of 
extinction corrections) into absolute magnitudes measured in Johnson B using

\setcounter{equation}{0}
\begin{equation}
M_B = m_Q - 5\log \left(\frac{D_L(z)}{10pc}\right) - K_{BQ}(z),
\label{EquRestFrameB}
\end{equation}

\noindent where $M_B$ is the predicted absolute magnitude of the object in B, 
$m_Q$ is the apparent magnitude of the object in some filter Q, $D_L(z)$ is the 
luminosity distance\footnote{All cosmological distances in this 
paper were calculated assuming a flat universe with $\Omega_m=0.27$ 
and $\Omega_\lambda=0.73$.} to the object at redshift $z$ and 
$K_{BQ}(z)$ is the K-correction from the object's redshift and filter Q to the 
rest-frame in B. We adopt the redshifts of \citet{JMK2004} for L1 and L2. The K-corrections were
calculated using the {\sc synphot} package in \iraf, together with template spectra
for L1 and L2 from \citet{BC1993} and \citet{Kinney1996}. 

\donewparg The K-correction for L1 was obtained from the model elliptical
galaxy spectra of \citet{BC1993}, who provide multiple templates with 
various star formation histories and initial mass functions (IMFs). 
We selected the template that produced the lowest scatter in the predicted 
rest-frame B magnitudes for L1 amongst the three filters (F555W, F814W and F160W). 
The optimum template gave an RMS scatter amongst the three filters of $\sim0.02$~mag, 
and was generated assuming a Salpeter IMF with mass range from 0.1 to 125 $M_\odot$, in 
which star formation took place $>$5~Gyr ago during a burst lasting 1~Gyr. The templates
were used solely to obtain a spectral K-correction; no correction for luminosity evolution
was applied.

\donewparg The K-correction for L2 was estimated using the starburst galaxy spectrum 
templates of \citet{Kinney1996}. Since there are several templates provided, each
having differing amounts of internal extinction, we again selected the template 
that produced the lowest scatter in the predicted rest-frame B magnitudes for L2, but between 
the F555W and F814W filters only. This optimum template had 0.15 $<$ E(B-V) $<$ 0.25 
and the resulting scatter was $\sim$0.01~mag. Again, no luminosity evolution correction was applied.

\donewparg Applying Equation \ref{EquRestFrameB} to the HST photometry we find a 
rest-frame magnitude $M_B = -20.57 + 5\log h \pm 0.1$ for L1, and $M_B = -13.95 + 5\log h \pm 0.1$
for L2, where we use $H_0 = 100 h $~\hunits. We compare this to the local characteristic absolute magnitude 
in B measured in the rest frame, $M_*$; \citet{Rusin2003} use lens galaxies
to find a value of $M_* = -19.7 + 5\log h \pm 0.29$, which implies a luminosity for L1 of $2.2\pm0.2 L_*$, 
and $5.0\pm0.5\times10^{-3} L_*$ for L2, where $L_*$ is the luminosity corresponding to $M_*$.

\donewparg The fundamental plane is defined by

\setcounter{equation}{1}
\begin{equation}
\log R_e = \alpha\log \sigma_c + \beta <\mu>_e + \gamma,
\end{equation}

\noindent in which $R_e$ is the effective radius (the radius containing half the total light) of
the galaxy in kpc, $\sigma_c$ is the central velocity dispersion of the luminous matter, $<\mu>_e$
is the galaxy's effective surface brightness in mag arcsec$^{-2}$, and $\alpha$, $\beta$ and $\gamma$
are empirically determined constants. The effective surface brightness is the mean surface brightness
within the effective radius, and is given by

\setcounter{equation}{2}
\begin{equation}
<\mu>_e = m_Q - K_{BQ}(z) + 2.5\log 2\pi + 5\log r_e - 10\log(1+z),
\end{equation}

\noindent in which $r_e$ is the effective radius in arcseconds, and $m_Q$ is the total magnitude of the object (i.e.
the magnitude within an infinite aperture). We have used the total magnitude derived from the S\'{e}rsic
fits, on the assumption that the extrapolation introduces minimal additional error compared to the 
uncertainties in K-corrections and magnitude system zero points. We do not apply an evolution correction to
$<\mu>_e$. We adopt the values for $\alpha$, $\beta$ and $\gamma$ found by \citet{LocalFP} for local ellipticals,
$\alpha=1.25$, $\beta=0.32$, $\gamma=-8.895$ for $H_0 = 50$~\hunits. The fundamental plane remains valid at 
redshifts comparable to that of L1 (eg \citealt{FarFP}). Assuming that L1 lies precisely on 
the fundamental plane, we use the results from the S\'{e}rsic fits to estimate $\sigma_c = 117^{+24}_{-20} h^{-0.8}$.

\donewparg We make an assumption of isothermality for the lens galaxy L1 in order to estimate the source redshift. 
Lens galaxies are typically found to have mass profiles within 15\% of isothermal within the radius of the lensed images 
\citep{Rusin2002,KT2003,RKK2003,TK2004,RK2004}. For an isothermal galaxy we can relate (dynamical) velocity dispersion 
to a function of the source redshift \citep{KSB}, because the velocity dispersion determines the angular separation 
of lensed images, $\Delta\theta$, through 

\setcounter{equation}{3}
\begin{equation}
\Delta\theta = 8\pi{\left(\frac{\sigma_{iso}}{c}\right)}^2 \frac{D_{ls}}{D_s},
\end{equation}

\noindent where $\sigma_{iso}$ is the velocity dispersion of the isothermal mass (not necessarily equal to $\sigma_c$), 
and \dds\ and \ds\ are the angular size distances from the lens to the source and from the observer to the source 
respectively. In Figure \ref{VelDispPlot} we have used this relation with the observed image splitting to plot the 
expected $\sigma_{iso}$ as a function of source redshift for $h=0.7$, assuming that $\sigma_c$ and $\sigma_{iso}$ are 
approximately equal \citep{CSKveldisp}. We also plot the two source redshifts suggested by \citet{JMK2004}, which 
between them conveniently span much of the plausible source redshift space. The estimated velocity dispersion of L1 
implies that the source redshift $z_s$ is greater than 1.2 at 95\% confidence; a source with $z_s = 0.8$ would require 
the lens to have a velocity dispersion of $\sim$330~kms$^{-1}$, assuming that L1 is isothermal. 

\setcounter{figure}{10}
\begin{figure}
\includegraphics[width=8cm]{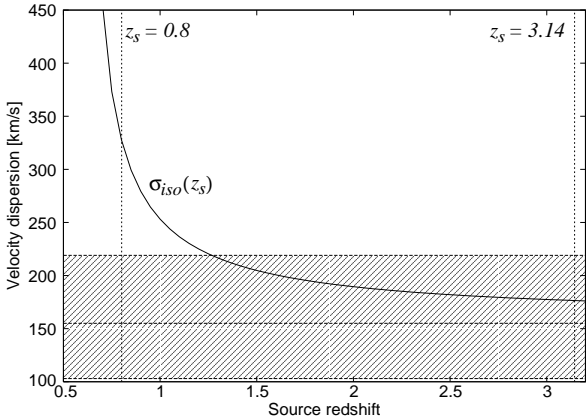}
\caption{Velocity dispersion of L1 ($\sigma_{iso}$) as a function of source redshift ($z_s$), assuming an 
isothermal mass distribution in the lens. The shaded area represents the velocity dispersion derived from 
the fundamental plane at the 95\% confidence level when $h=0.7$. Assuming an isothermal mass distribution, we find 
$z_s>1.2$ with 95\% confidence.}
\label{VelDispPlot} 
\end{figure}

\donewparg The data available favour a high source redshift, but further spectroscopy will be 
essential if the radio source proves to be time-variable, since then B0631+519 would be a potential target for 
attempts to measure the Hubble constant. A spectroscopic source redshift would also be valuable for 
normalising mass models of this system. 

\section{Lens modelling of B0631+519}
\label{SecModel}

In this Section we seek to demonstrate that the lens hypothesis is a 
plausible explanation for the radio emission seen from B0631+519, by 
comparing the predictions of simple parametric mass models to the observations. 
We fit mass models to the 5~GHz MERLIN data on A1, A2, B1 and B2 
because these images are relatively compact. We then take the optimum 
mass model fitted to the radio data and use it to invert
the Einstein ring seen in the NICMOS F160W image by applying semi-linear inversion
techniques \citep{WD2003,TK2004,K2005}. Finally, we use the mass model to
examine L1's mass to light ratio. The unknown source redshift means that the 
estimated time delays, as well as L1's mass (within the Einstein radius) and 
its mass-to-light ratio are scaled by factors that depend on $z_s$ through 
combinations of the angular diameter distances between source, lens and observer.
We therefore quote the affected quantities for $z_s=0.80$ and $z_s=3.14$.

\subsection{Mass modelling using the 5~GHz MERLIN data}

When fitting a mass model to the radio observations, the observed image positions 
are back-projected to the source plane through the current mass model. These back-projected
images are used to update positions and flux densities of the model sources, and the updated 
sources are then projected forward onto the image plane and compared to the observed image 
positions and flux densities through a standard image-plane $\chi^2$ statistic \citep{CSK91}. 
The statistic is defined over N images having angular positions $\btheta_i$ 
($i=1...N$) and flux densities $s_i$ by

\setcounter{equation}{4}
\begin{equation}
\chi^2 = \sum_{i=1}^{N} \frac{|{\btheta_i - \btheta_i'}|^2}{{\sigma_{pi}}^2} + \sum_{i=1}^{N}\frac{|{s_i - s_i'}|^2}{{\sigma_{fi}}^2},
\end{equation}

\noindent where the primed quantities are values predicted by the lens model being 
evaluated and the $\sigma_{pi}$ and $\sigma_{fi}$ quantities account for errors in positions
and flux densities respectively. The lens model parameters are iteratively optimised by a downhill 
simplex method \citep{PressEtAl} until a minimum is found. In this approach we assume
that the observed images are unresolved, which is clearly the case for A1 and B1 in the
MERLIN 5~GHz data. A2 and B2 are slightly resolved but are still well-represented by
Gaussian models. Components A1a, A1b, A1c and A1d will become useful if future 
high-resolution VLBI observations are able to resolve B1 into corresponding 
images. Without observed counterparts in B, the VLBA sub-components in A1 
cannot yet be used to constrain the lens model. Images X, Y and Z are very
extended and would be best treated using tools such as \lensmem\ \citep{LensMEM} 
or \lensclean\ \citep{LensCLEAN1,LensCLEAN2}.

\donewparg Because A1 and B1 do not arise from the same source as A2 and B2, this
system has fewer degrees of freedom immediately available than for a 
single quadruply imaged source. We ignore any contribution L2 
might make to the lens effect and treat L1 as a singular isothermal ellipsoid 
(SIE; \citealt{KSB}). We allow the lens position to vary freely. The constraints 
available for this simple lens model are then four sets of image positions 
and flux densities. The free parameters are the positions and flux densities 
for two sources, the lens position, the lens mass scale and the position-angle 
and ellipticity of the mass distribution. The number of degrees of freedom, $\nu$, is 
therefore 1. We allow 20\% errors on image flux densities to account for
possible source variability, and we use the positions and position
errors derived from the MERLIN 5~GHz data for A1, A2, B1 and B2 (because the 
VLBA observations resolve A2 and B2). The model parameters are shown in 
Table \ref{ModelsT}, and the lens is illustrated in Figure \ref{ModelsF}.
The optimum model has a reduced $\chi^2$ of 0.41, with roughly equal 
contributions from flux densities and image positions. 

\setcounter{table}{7}
\begin{table}
\begin{center}
\begin{tabular}{llll}
\hline
Model & SIE & SIE$+$shear \\
\hline
Lens offset in R.A. [mas] & $+$398$\pm$23 & $+$370 (fixed)\\
Lens offset in Dec. [mas] & $-$519$\pm$5 & $-$520 (fixed)\\
Critical radius [mas] & 603$\pm$2 & 605$\pm$3 \\
Ellipticity & 0.13$\pm$0.03 & 0.07$\pm$0.04  \\
Lens P.A. & $-$57.1$\pm$0.1\degr & $-$28$\pm$12\degr \\
Shear strength &  & 0.03$\pm$0.01 \\
Shear P.A. &  & $-$81$\pm$0.5\degr \\
$\chi^2/\nu$ & 0.41/1 & 0.07/1 \\
\hline
\end{tabular}
\caption{Lens models for B0631+519. Lens position offsets are given
with respect to image A1. Position angles are measured east of north. The errors in the 
lens parameters were estimated using Monte Carlo simulations.}
\label{ModelsT}
\end{center}
\end{table}

\donewparg Next, we fixed the galaxy position to that determined optically in Section 
\ref{SecOptObs}, and added an external shear to the model. This 
SIE+shear model was then fitted to the MERLIN image positions and flux densities. The
optimum model had a reduced $\chi^2$ of only 0.07 for 1 degree of freedom,
suggesting that the random errors on the constraints were over-estimated.
Reducing the flux density errors from 20\% to 5\% results in a reduced $\chi^2$
of 0.5 which is dominated by the contributions from the image positions. The 
optimum model parameters do not change significantly when the flux density 
errors are reduced. The optimum model parameters are listed in Table 
\ref{ModelsT} and the model is illustrated in Figure \ref{ModelsF}. We note that
the position of the best SIE-only lens is consistent with the optical lens 
position within the errors. The time delay between A1 and B1 is predicted to be 
10.2 $h^{-1}$ days with the SIE model or 8.5 $h^{-1}$ days with the SIE+shear model
for a source redshift of 3.14.

\setcounter{figure}{11}
\begin{figure}
\includegraphics[width=7cm]{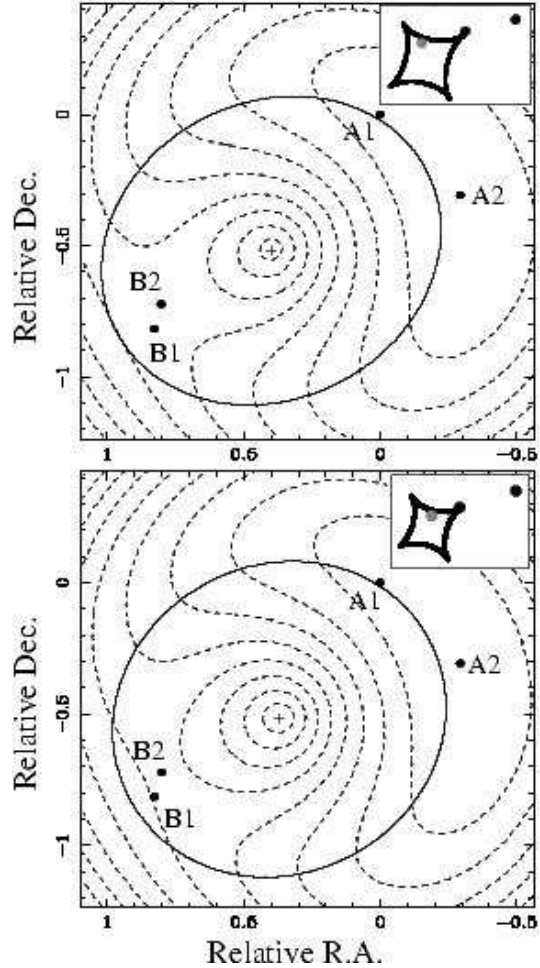}
\caption{Lens models applied to B0631+519. The upper panel shows the best single SIE model. 
The lower panel shows the best SIE+shear model when the lens galaxy position is fixed to 
the measured optical/infrared galaxy position. The ellipse in each panel is the 4-image critical 
curve. The dashed contour lines represent the time delay surface of the lens relative
to image A1. The contours in the upper panel begin at 1.23 $h^{-1}$ days (for the first contour 
surrounding A1/A2) and increase in steps of 2.05 $h^{-1}$ days. In the lower panel, the first
contour surrounding A1/A2 represents 1.11 $h^{-1}$ days and the contours increase in steps of 
1.88 $h^{-1}$ days. The contour levels are calculated for a source redshift of 3.14, and should
be multiplied by a factor of 3.5 for $z_s=0.80$. The lensed images are represented by filled 
circles. The inset diagram shows the source plane; the astroid curve is the 4-image caustic, and 
filled circles represent the modelled source positions. The estimated position of the source 
responsible for images Y, Z and the arc X is marked by a light grey filled circle. }
\label{ModelsF}
\end{figure}

\donewparg The lens modelling we have performed shows that the A1, A2, B1 and B2
components can be understood as lensed images of two distinct sources.
It seems that A1 and B1 are images of a compact radio core having a flat 
(synchrotron self-absorbed) spectrum, while A2 and B2 arise from a more 
extended region of emission to the west of the core, probably a radio lobe. 
Images A1 and A2 have positive parity, while images B1 and B2 have negative parity.
Incorporating images X, Y and Z is more difficult because they are
all relatively extended compared to the A and B images, and the radio spectra
indicate that these images arise from a third source. Back-projecting
the positions of the Y and Z images (from the 1.7~GHz MERLIN data) to 
the source plane, along with the brightest point on arc X, produces an
approximate source position of $+$0.41\arcsec east and $-$0.49\arcsec north relative 
to A1 (marked on Figure \ref{ModelsF}). The 
model thus suggests that the X, Y and Z images are produced by a 
counter-lobe on the eastern side of the source that happens to lie 
inside the 4-image caustic. With only the VLA data, showing X and Y 
but no Z, a model involving emission only on the western side of the 
source's core might also be plausible. 

\subsection{Non-parametric de-lensing of the Einstein ring}

The blue color of galaxy L2 and its consequent apparent absence in the HST-NICMOS 
F160W image, allows us to make use of the infrared Einstein ring to further 
test the validity of the lens mass model of CLASS 0631+519, which is based 
on the MERLIN radio data alone.

\donewparg The Einstein ring is de-lensed (see \citealt{WD2003,TK2004,K2005}), 
using the SIE plus shear model from
Table \ref{ModelsT}. It is stressed that the lens mass model is {\sl not} adjusted
in any way to better fit the observed Einstein ring. A TinyTim
generated PSF is used to account for blurring of the image. The
structure of the Einstein ring is remarkably well reproduced (Figure \ref{DeLensed}),
given that the lens mass model was only fitted to the radio jet
structure.  Despite the overall good quality of the reconstruction, a
clear deviation from the model is found in the residual image, just
north of L1 on the Einstein ring and slightly offset from the position
of L2.  Neither varying the mass model, nor alleviating the
regularization can remove the feature completely.

\donewparg We propose two possible explanations for this apparent
``anomaly'': (a) it is due to residual emission from L2 -- still
present at a low level in the NICMOS-F160W image -- or a dwarf-galaxy
associated with L1, or (b) it is due to a small-scale perturbation of
the lens mass model by the presence of L2, creating an anomaly in the
ring \citep{K2005} similar to flux-ratio anomalies in compact lensed
sources (e.g.\ \citealt{MS1998}). Although the first explanation(s) 
can not be completely excluded, the anomaly is situated near the brightness peak of the
Einstein ring, appears to follow the curvature of the ring and is
offset from the centroid of L2. Each of these observations seem to
support the second explanation. As shown in \citet{K2005}, one would 
indeed expect such anomalies in Einstein rings, as a result of tiny 
perturbations of the dominant lens potential. The foreground galaxy L2 
could provide the required perturbation. A full analysis of the nature 
of the anomaly, however, is beyond the scope of this paper.

\setcounter{figure}{12}
\begin{figure}
\includegraphics[width=8cm]{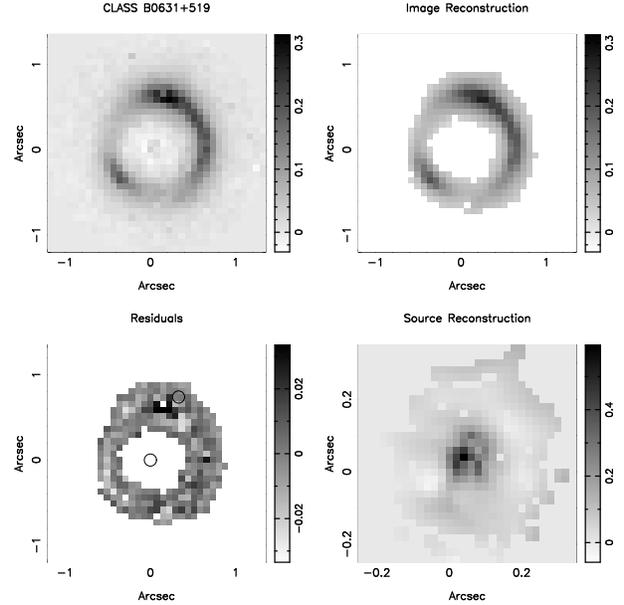}
\caption{Non-parametric lens reconstruction of CLASS B0631+519. {\sl
Top-left:} The galaxy-subtracted NICMOS image of the
Einstein ring. {\sl Top-right:} The best reconstruction of the
observed Einstein ring.  {\sl Bottom-left:} The residuals between
observations and the model.  {\sl Bottom-right:} The reconstructed
source brightness distribution. For the source reconstruction, we use
a $30\times 30$ pixel grid with a scale of 0.02 arcsec~pix$^{-1}$. We
mask all image pixels below 3\,$\sigma$ (indicated in the image
reconstruction; $\sigma=0.011$~counts~pix$^{-1}$). The regularization
parameter for the source is $\lambda_{\rm s}=0.04$, resulting in
$\chi^2/{\rm d.o.f.}\approx 1$.  We use a regularization matrix that
minimizes the curvature of the source (see \citealt{K2005}). Note that a
significant ``anomaly'' is left in the residual image (see text).  The
two circles in the residual image indicate the centroids of L1 (at the 
origin of the plot) and L2 at $-$0.33~arcsec east, $+$0.74~arcsec north of L1}
\label{DeLensed}
\end{figure}

\subsection{Mass to light ratio of L1}

\donewparg We have sufficient information to estimate the B-band 
mass-to-light ratio of L1 in the region enclosed by the lensed images. 
The mass $M$ between the lensed images of a source (in the case of a circularly symmetric lens) 
is related to the image separation $\Delta\theta$
and the angular diameter distances $D_l$ (observer to lens), $D_s$ 
(observer to source) and $D_{ls}$ (lens to source) by 

\setcounter{equation}{5}
\begin{equation}
\label{EinMass}
M = 3.074\times10^{7}~\frac{D_l D_s}{D_{ls}}~{\left( \Delta\theta \right)}^2,
\end{equation}

\noindent where $\Delta\theta$ is given in arcseconds. The estimate is significantly affected by the unknown source redshift
(but not by assumptions concerning the distribution of mass inside the radius of the images).
The ratio of angular diameter distances in Equation \ref{EinMass} decreases by
a factor of $\sim$3.5 when the source redshift is increased from 0.8 to 3.14.
We therefore estimate the lensing mass using both of the suggested redshifts for 
the source listed in \citet{JMK2004}, since they conveniently span most of the range 
of plausible source redshifts. The separation between images A1 and B1 is 1.162$\pm$0.003~arcsec, 
which at the redshift of L1 corresponds to a linear distance of $5.6h^{-1}$~kpc. 
From Equation \ref{EinMass} we estimate the mass of L1 within the images at 
$2.17\pm0.01\times10^{11} h^{-1}$ $M_\odot$ using a source redshift of $z_s=0.80$ or 
$6.27\pm0.03\times10^{10} h^{-1}$ $M_\odot$ using $z_s=3.14$. Combining these estimates 
with the rest frame B-band absolute magnitudes derived from the S\'{e}rsic profile fits in Section \ref{SecRedshifts} 
(corrected for a 1.16~arcsec diameter aperture) produces estimates for the mass-to-light ratio of L1 in 
the B filter of $15\pm2 h$ $M_\odot/L_\odot$ using $z_s=0.80$, or $4.3\pm0.4 h$ $M_\odot/L_\odot$ using $z_s=3.14$.
We have taken the absolute magnitude of the Sun in B to be 5.54. The mass-to-light ratio estimates 
apply to the central region of L1 encircled by the lensed images (i.e. that within a mean radius 
of 2.8~$h^{-1}$~kpc from L1's centre). Both figures are comparable with $M/L$ estimates from local 
ellipticals \citep{Gerhard}, although the figure for $z_s=3.14$ is relatively low given $h=0.7$.

\section{Conclusions}
\label{SecConclude}

Together the radio and optical/near-infrared observations presented here show
that CLASS B0631+519 can be understood as a gravitational lens system. A 
simple SIE+shear mass model successfully reproduces features of the radio 
observations, and a non-parametric approach to the near-infrared data has 
identified the perturbing effect of L2 on the smooth lensing properties of the 
system. The mass modelling done so far leads us to conclude that the background 
radio source probably consists of three radio components in a classic double-lobed 
core-jet arrangement. The eastern lobe is located inside the quad-image caustic 
for the system and is lensed into the arc X and the more compact images Y and Z. 
The core region is doubly imaged into A1 and B1, while the western lobe is lensed into
images A2 and B2. The magnification provided by the lens and the high 
resolution of the VLBA has allowed identification of the flat-spectrum AGN core 
and revealed jet components from both the approaching and the receding jets. 
The near-infrared source centre coincides with the radio source positions to
within the errors.

\donewparg The wealth of constraints available means that this system will 
benefit from application of deconvolution/inversion techniques that can utilise
the extended images seen in the radio data. Having a good mass model for this 
system will allow us insights into the mass distributions of ellipticals at 
redshifts of 0.6, information that is not readily available in any other way. 
The hints of radio source variability mean that it may be possible to determine 
a radio time delay and use it to measure the Hubble constant.

\donewparg This lens system is relatively unusual in having two galaxies of different
redshifts present along the line of sight to the source, a situation that also occurs
in CLASS B2114+022 \citep{2114_1,2114_2} and in PKS 1830-211 \citep{1830a}. In B2114+022
the effect of the nearer lens is much more significant than in B0631+519 \citep{2114_3}.
In PKS 1830-211 the nearer lens was detected through its HI absorption \citep{1830b} at
a redshift of 0.19 (the main lensing galaxy has a redshift of 0.886). The correct 
interpretation of the optical data in PKS 1830-211 is presently unclear, although the 
main lens is certainly a face-on spiral \citep{1830c,1830d}.

\donewparg Although L2 appears to have a relatively minor influence on the lensing properties of B0631+519, 
the radio emission from the lensed source must pass through L2's ISM. Searches for 
molecular absorption lines may be worthwhile to provide extra information on the 
nature of this galaxy. The mass distribution in L2 may well be constrainable
with further non-parametric modelling work.

\section*{Acknowledgments}

The authors would like to thank Shude Mao for useful discussions,
and the referee, Stephen Warren, for comments that improved the clarity of the paper.
MERLIN is a UK National Facility operated by the University of
Manchester on behalf of PPARC. The VLA and VLBA are operated by the
National Radio Astronomy Observatory for Associated Universities
Inc. under a co-operative agreement with the National Science
Foundation. This research used observations with the Hubble Space
Telescope, obtained at the Space Telescope Science Institute, which is
operated by Associated Universities for Research in Astronomy, Inc,
under NASA contract NAS 5-26555; these observations are associated
with HST programme 9744. The William Herschel Telescope is operated 
on the island of La Palma by the Isaac Newton Group in the Spanish 
Observatorio del Roque de los Muchachos of the Instituto de 
Astrofisica de Canarias. This research has made use of the NASA/IPAC
extragalactic data base (NED) which is operated by the Jet Propulsion
Laboratory, Caltech, under contract with the National Aeronautics and
Space Administration. This publication makes use of data products from
the Two Micron All Sky Survey, which is a joint project of the
University of Massachusetts and the Infrared Processing and Analysis
Center/California Institute of Technology, funded by the National
Aeronautics and Space Administration and the National Science
Foundation. This work was supported by the European Community's Sixth 
Framework Marie Curie Research Training Network Programme, Contract No. 
MRTN-CT-2004-505183 ``ANGLES''. RDB is supported by NSF grants AST-0206286 
and AST-0444059. TY, JPM, MAN and PMP acknowledge the receipt of PPARC 
studentships.

\label{lastpage}

\end{document}